\documentstyle[prl,aps,preprint,tighten]{revtex}
\begin{document}

\draft

\title{Inertial Mass and Viscosity
of Tilted Vortex Lines in Layered Superconductors}

\author{ A. S. Mel'nikov}
\address{Institute for Physics of Microstructures,
        Russian Academy of Sciences\\
         603600, Nizhny Novgorod, GSP-105, Russia}

\date{\today}
\maketitle
\begin{abstract}
The dynamics of tilted vortex lines in
Josephson-coupled layered superconductors is
considered within the time-dependent
Ginzburg-Landau theory.
The frequency and angular dependences of the
complex-valued vortex mobility $\mu$ are studied.
The components of the viscosity and inertial mass
tensors are found
to increase essentially for magnetic field orientations
close to the
layers. For superconducting/normal metal multilayers the
frequency ($\omega$) range is shown to exist where
the $\mu^{-1}$ value depends logarithmically on $\omega$.
\end{abstract}
\narrowtext
The inertial mass $M$ and viscosity $\eta$
are known to be important quantities for
characterizing vortex dynamic response in
type-II superconductors.
Without taking account of pinning effects, the vortex
line bending
and traction by the superflow
the simple phenomenological description of the vortex
dynamics in isotropic superconductors
(see, e.g. Ref.~\cite{gitt}) is based
on the following equation for the
flux line velocity ${\bf V}$:
${M\dot{\bf V} +\eta{\bf V}=\phi_0{\bf j}_{tr}\times{\bf n}/c}$,
where ${{\bf j}_{tr}={\bf j}_0exp(i\omega t)}$ is
the transport current density,
${\bf n}$ is the unit vector which points in the magnetic
field direction, $\phi_{0}$ is the flux quantum.
In the context of many applications (including
high-frequency phenomena such as microwave or
infrared response) it is convenient to introduce
also the complex-valued dynamic mobility
$\mu=(i\omega M +\eta)^{-1}$ which may be extended
to include both pinning and flux-creep effects
\cite{mobil,clem2}.
For $T$ close to $T_{c}$ the quantity
$\eta$ may be estimated within
the Bardeen-Stephen model or calculated more exactly
using the time-dependent Ginzburg-Landau (TDGL) theory
\cite{gorkov}.
The inertial effects may play an essential role
not only in high-frequency classical dynamics
but also in quantum tunneling phenomena at
extremely low temperatures \cite{review}.
There exist several contributions to the vortex inertial mass
\cite{review,suhl,kupr,duan2,duan,sim1,sim2,coffey},
i.e., one from the electronic states within the normal core
(the electronic contribution $M_c$),
one connected with the electric field
energy ($M_{em}$) and one arising from mechanical stress and
strain effects. Note that for isotropic compounds
the term $M_{em}$ is much less than the electronic
contribution ${M_c\sim \hbar^2 N_f}$
($N_f$ is the density of states at the Fermi level).

Recently a great deal of attention has been devoted to the
peculiarities of the vortex dynamics in
extremely anisotropic high-$T_{c}$ superconductors
which pertain to a larger class of materials including
quasi-two dimensional (2D) organic superconductors,
 chalcogenides and
artificial superconducting multilayers.
A common feature of these
systems is the weak interlayer Josephson coupling which
results in a short effective  coherence length
$\xi_{c}$ for the order parameter spatial
variation along the  $c$
direction (perpendicular to the layers).
In a broad temperature range
the $\xi_{c}$ value may be much smaller than
the interlayer distance
$D$. This fact results in the quasi-2D character of static and
dynamic magnetic properties in these compounds.
As a consequence, the description
of the dissipative and inertial effects in the vortex motion
within the anisotropic TDGL theory
employing an effective mass tensor
\cite{asm,hao,hao2,ivlev1} appears to be adequate
only over a limited temperature range
(${\xi_c(T)>D}$).
In this paper we restrict ourselves
to the case of weak magnetic fields
when the so-called effective cores
of neighbouring flux lines
do not overlap (see Refs.~\cite{feinberg,lnb3})
and study the motion of an isolated vortex line
which  forms the angle $\gamma$ with the $z$ axis
($c$ direction) and lies, e.g., in the $(xz)$ plane.
It is the purpose of this Letter to report on
a calculation of the vortex dynamic mobility
for layered superconducting structures
in the temperature range $\xi_c(T)\ll D$.
Let us consider a stack of thin superconducting (S) films
of thickness $d$ separated by insulating (I)
or normal metal (N) layers of thickness ${D\gg d}$.
The inertial effects and a viscous drag force acting
on a moving flux line are determined by
the complicated structure of the effective vortex core
\cite{feinberg,lnb3}.
For ${tan\gamma < \xi/D}$
the effective core size is of
the order of the coherence length $\xi$
in the plane $(xy)$.
In the angular domain ${Dtan\gamma > \xi}$
the effective core consists of 2D core regions
connected by the Josephson-like vortex cores
which appear for ${Dtan\gamma > L_{j}}$, where
$
L_{j}=(D\hbar c^{2}/
(8\pi e \lambda_{ab}^{2}J_{c}))^{1/2}
$
is the Josephson length, $J_c$ is the interlayer
Josephson critical current density, $\lambda_{ab}$
is the London penetration depth for currents along layers.
The characteristic dimension
of the 2D core region in the plane $(xy)$ is
$a_d=min[Dtan\gamma,L_j]$.
The dimensions of the Josephson-like core along
$y$ and $z$ are $L_j$ and $D$, respectively.

Before presenting the derivation of
the vortex dynamic equation, we
briefly discuss the main mechanisms
responsible for the inertial and
dissipative effects in the vortex motion.
One may expect that the contributions arising
from the regions of 2D pancake normal cores do not
change essentially and may be calculated in a manner
analogous to the
treatment in Refs.\cite{suhl,kupr,duan2,duan,sim1}.
The novel specific terms in $M$ and $\eta$ come from the
2D and Josephson-like core regions.
In particular, the simple qualitative arguments
show that with an increase in $\gamma$
the term $M_{em}$ should increase
 due to the capacitive effects.
This term is determined by the spatial distribution
of the electric field generated by a moving flux line.
Far away from the normal core regions and
for rather large distances between the neighbouring
2D pancake vortices (${Dtan\gamma > \xi}$)
the averaged $z$ component of the
electric field $E_{zn}$  between layers $n$ and ${n+1}$
may be found using the Josephson relation
(we take here the gauge ${A_z=0}$):
${E_{zn}=-\hbar {\bf V}\nabla (\theta_n-\theta_{n+1})/(2eD)}$,
where ${\theta_n ({\bf r}-{\bf R}(t))}$ is the phase of
the complex order parameter in the layer $n$,
${\dot{\bf R}={\bf V}}$ is the vortex velocity.
At large distances $\tilde R_n$ from the 2D vortex center
(in the 2D core region)
${E_{zn}}$ decreases as  $\tilde R_n^{-1}$ to the distances
${\tilde R_n\sim a_d}$.
Such an extremely slow decay of ${E_{zn}}$
results in the logarithmic divergence
of the kinetic energy of the electric field,
which is cut off at a large length scale $a_d$.
Thus the contribution proportional to the value
${ln\,(a_d/\xi)}$ should appear in the expression for the
inertial mass. This enhancement of the $M_{em}$ value may
be significant if the dielectric constant of insulating
layers is large. The analogous logarithmic term from the
2D core region contributes to the
viscosity coefficient due to the
dissipation produced by the interlayer
normal currents (see also
Ref.~\cite{mel}). Surely, the influence
of these interlayer currents
on the scalar electrodynamic potential distribution
(and, hence, on the complex-valued dynamic mobility)
is most significant for S/N multilayers and
temperatures close to $T_c$.
For large angles ${tan\gamma > L_j/D}$
the formation of Josephson-like cores results in
the additional contributions to the vortex viscosity and mass,
which should be linear in the length $Dtan\gamma$.
These terms may play an essential role in the vortex dynamics
as it follows from the results obtained
in Refs.~\cite{clem2,clem1}
for the particular case ${\gamma=\pi/2}$.
Obviously, the Josephson strings parallel
to the $x$ axis contribute to
the vortex mobility only for a nonzero velocity component $V_y$.
As a consequence, for large tilting angles
the vortex mass and viscosity
coefficient should depend on the transport current orientation
in the plane $(xy)$.

We now continue with the calculation of the viscosity
coefficient and
the vortex mass within the
phenomenological Lawrence-Doniach model \cite{ld}
generalized for the
description of the dynamic phenomena
(for $T$ close to $T_{c}$).
Such a generalization may be written in analogy to
the TDGL theory in 3D homogeneous
superconductors. In the following treatment
we will use the diffusion type
TDGL equations which are strictly valid only for
gapless superconductors.
Nevertheless, we believe the final results to
be qualitatively correct for a more general case.
Note also that in the TDGL equations given below
we neglect the terms connected with the particle-hole
asymmetry and responsible for the vortex traction by the
superflow \cite{dorsey,kopnin}.
 If we choose the gauge
$A_z=0$, the equation for the order parameter
$\psi_n({\bf r})={f_{n}exp(i \theta_{n})}$
and the current density (averaged over the
periodicity length $D$) have the form:
\begin{eqnarray}
\tau\left(\frac{\partial}{\partial t}+
\frac{2ie\varphi_n}{\hbar}
\right)\psi_n=
\xi^{2}\left(\nabla-\frac{2ie}{\hbar c}{\bf A}_n \right)^{2}
\psi_{n}+\psi_{n}\nonumber\\
\label{main1}
-|\psi_{n}|^{2}\psi_n
+\frac{\xi^2}{L_j^{2}}(\psi_{n+1}+\psi_{n-1}-2\psi_{n})
\\
\label{main2}
{\bf j}_{n\parallel}=
\frac{\hbar c^{2}f_{n}^{2}}{8\pi e \lambda_{ab}^{2}}
(\nabla\theta_{n}-
\frac{2e}{\hbar c} {\bf A}_{n})
-\sigma_{ab}
(\nabla \varphi_n+
\frac{1}{c}\dot{\bf A}_n),
\end{eqnarray}
Here $\varphi_{n}$
is the electrochemical potential in the $n$-th layer,
$\sigma_{ab}$ is
the normal state conductivity in the layer direction,
${\tau=\pi\hbar/(8(T_{c}-T))}$.
The interlayer current density is given by the expression:
\begin{equation}
\label{jz}
(j_{z})_{n,n+1}=J_c f_n f_{n+1}
sin(\tilde\theta_{n+1,n})-
\sigma_c D^{-1}\tilde\varphi_{n+1,n}
\end{equation}
where $\sigma_c$ is the interlayer normal state conductivity,
$\tilde\theta_{n+1,n}= \theta_{n+1}-\theta_n$,
$\tilde\varphi_{n+1,n}=\varphi_{n+1}-\varphi_n$.
Employing the expressions (\ref{main2}),(\ref{jz})
for S/N multilayers
we assume that all the relevant length scales
in the plane $(xy)$ are larger than the periodicity
length $D$. As a consequence, our approach is strictly valid
only in the limit $\xi\gg D$ \cite{mel}.
We consider the dynamics of a tilted vortex line
in the presence of an applied transport current
${\bf J}_{n}(t)$ flowing along the plane $(xy)$.
For simplicity we suggest that
the total stack thickness is much
smaller than the effective skin depth and
the current density ${{\bf J}_{n}={\bf J}_{tr}}$
does not depend on the layer number $n$.
Considering only straight tilted vortices,
we also leave aside the problem of possible vortex
line flexures immediately under the surface.
According to Ref.~\cite{brandt}, for the particular case
$Dtan\gamma\ll L_j$  these flexures occur in
the boundary layer of the width
${L_s\sim min[\lambda_{ab},a] D/L_j\ll\lambda_{ab}}$,
where $a$ is the intervortex spacing.
The generalization of the technique discussed
below, taking account of the vortex line bending is
straightforward (see, e.g., Ref.~\cite{gorkov}).

To obtain the equation of motion we use the procedure
similar to the one discussed in Refs.~\cite{gorkov,mel,dorsey}
and search for the solution
of Eq.(\ref{main1}) in the form:
\begin{equation}
\label{pert}
{\psi_{n}=G_n+g_n=G({\bf r}-{\bf R}-{\bf r}_{n})+
g({\bf r}-{\bf R}-{\bf r}_{n},t)},
\end{equation}
where ${{\bf r}_n=(nDtg\gamma,0)}$,
$G({\bf r}-{\bf r}_{n})$ is the order
parameter distribution for a static vortex line and $g$ is
the small correction which is of the order of
${{\bf V}=\dot{\bf R}}$.
The arguments analogous to the ones used for
isotropic type-II superconductors \cite{gorkov}
show that in the limit $\lambda_{ab}\gg D$
the magnetic field generated by
a moving vortex line may be neglected
within the effective core region
(which provides the main contribution
to the inertial mass and viscosity).
The condition $\lambda_{ab}\gg D$ may be
easily met in real layered structures,
at least, for temperatures
close to $T_c$. As a consequence, in
Eqs.~(\ref{main1}),(\ref{main2})
we may take into account only the vector potential
${{\bf A}_n=-4\pi\lambda_{ab}^2c^{-1}{\bf j}_{s}}$
generated by the transport supercurrent
density ${{\bf j}_{s}}$ averaged over the structure period.
Substituting the expansion
(\ref{pert}) into Eq.~(\ref{main1}) and
neglecting the higher order
terms in $V$ one obtains the equation
for the correction $g_n$.
We next multiply this equation
 by ${G_{pn}={\bf p}\nabla G_n}$
(${\bf p}$ is an arbitrary translation
vector in the plane $(xy)$)
and integrate over the area $S$  of the circle
${|{\bf r}-{\bf r}_n-{\bf R}|\leq R_{1}}$
where $R_{1}$ meets the condition
 ${R_{1}\gg Dtan\gamma}$.
After simple transformations
the real part of the resulting equation reads:
\begin{equation}
\label{equ1}
\frac{\phi_{0}}{c}{\bf p} ({\bf j}_{s}\times{\bf z}_0)=
\frac{\eta_0}{\pi}Re\int\limits_{S}
G_{pn}^* (G_{vn}  -
\frac{2ie}{\hbar}\varphi_n G_n-
\dot g_n) d^2 r,
\end{equation}
where
$\eta_{0}=0.5\sigma_{ab} u\phi_{0} H_{c2}c^{-2}$,
$H_{c2}$ is the upper critical field for ${\gamma = 0}$,
${u=\hbar c^{2}/(32\lambda_{ab}^{2}\sigma_{ab}(T_{c}-T))}$,
and ${G_{vn}={\bf V}\nabla G_n}$.
The numerical factor $u$ is determined by the pair-breaking
mechanism (see Refs.~\cite{watts,eliash} for details).
The term in the left-hand side (l.h.s.)
of Eq.~(\ref{equ1}) corresponds
to the Lorentz force acting on a 2D pancake vortex while
the right-hand side (r.h.s.) contains the terms responsible for
the viscosity and inertial mass.
We do not consider the extrinsic
forces due to pinning and the interaction with
other vortex lines, which add to the l.h.s.
of Eq.~(\ref{equ1}).
Let us examine at first the contribution to the integral in the
r.h.s. of Eq.~(\ref{equ1}) which comes from the domain
${|{\bf r}-{\bf R}-{\bf r}_{n}|\gg\xi}$ (outside
the normal cores). In this case one can put
${|\psi_{n}|\simeq 1}$.
The continuity equation for the layered system reads:
\begin{eqnarray}
\label{cont1}
\frac{\partial \rho_n}{\partial t} + div{\bf j}_n
+\frac{(j_z)_{n,n+1}-(j_z)_{n-1,n}}{D} = 0,\\
\label{cont2}
\rho_n = \frac{\varepsilon}{4\pi D^2}
(2\varphi_n - \varphi_{n+1}-\varphi_{n-1}),
\end{eqnarray}
where $\rho_n$
is the averaged charge density in the layer $n$ and
$\varepsilon$ is the high-frequency dielectric constant.
We neglect here the difference between the electrochemical and
scalar electrodynamic potentials, i.e., assume
the Thomas-Fermi screening length to be less than
all the relevant length scales.
Using Eqs.~(\ref{main1}),(\ref{cont1}),(\ref{cont2}) one obtains:
\begin{eqnarray}
\label{phase}
\frac{2e\tau}{\hbar}\Phi_n=
\xi^2\Delta\theta_n + \frac{\xi^2}{L_j^2}
(sin\tilde\theta_{n+1,n}-sin\tilde\theta_{n,n-1})\\
\label{poten}
\Phi_n=\left(s+\frac{\varepsilon s}{4\pi\sigma_c}
\frac{\partial}{\partial t}\right)
(\varphi_{n+1}+\varphi_{n-1}-2\varphi_{n}),
\end{eqnarray}
where ${s=\sigma_c\xi^2/(u\sigma_{ab} D^2)}$
and
$
{\Phi_n=\varphi_n+\hbar\dot\theta_n/(2e)}
$
is the gauge-invariant scalar potential.
Note that we assumed ${u\stackrel{_>}{_\sim}1}$ which is
consistent with the microscopic theory results for gapless
superconductors \cite{watts,eliash}.
Let us consider the two-term
expansion for the order parameter phase:
${\theta_n=\theta_{nv}+\chi_n
=\theta_{v}({\bf r}-{\bf R}-{\bf r}_n)
+\chi_n}$,
where $\theta_v({\bf r}-{\bf r}_n)$ is the phase distribution
for a static vortex line and $\chi_n$ is the first-order in
$V$ correction.
To solve the system (\ref{phase}),(\ref{poten})
 we use here the linear approximation for the
interlayer Josephson current density  and replace the terms
${sin\tilde\theta_{n+1,n}}$ by ${\tilde\theta_{n+1,n}}$.
The validity of
such an approximation was discussed in detail
in Ref.\cite{lnb3}.
The solution of the system (\ref{phase}),(\ref{poten})
 gives us a possibility to evaluate
the integral in the r.h.s. of Eq.~(\ref{equ1})
far away from the normal cores
taking account of the expressions
${G_{pn}\simeq i({\bf p}\nabla\theta_{nv})exp(i\theta_{nv})}$,
${g_n\simeq i\chi_n exp(i\theta_{nv})}$.
Note that the contribution which comes from
the normal core region
(${|{\bf r}-{\bf r}_n-{\bf R}|\stackrel{_<}{_\sim}\xi}$)
does not differ essentially from the one obtained previously
for a vortex line in isotropic
superconductors \cite{gorkov,kupr}.
Finally one obtains the following equation of
vortex line motion in the Fourier representation:
\begin{equation}
\label{mot}
\mu_x^{-1} V_x(\omega){\bf x}_0 +
\mu_y^{-1} V_y(\omega){\bf y}_0 =
\frac{\phi_{0}}{c}{\bf J}_{tr}(\omega)\times{\bf z}_0,
\end{equation}
where ${\bf x}_0$, ${\bf y}_0$, ${\bf z}_0$
are the unit vectors of the coordinate system.
We follow here the treatment in Refs.~\cite{gitt,kupr}
and include both the supercurrent and normal current densities
to the driving Lorentz force in the r.h.s. of
Eq.~(\ref{mot}).
The complex-valued dynamic mobilities $\mu_{x,y}$
appear to depend strongly on the
parameter $s$ and tilting angle $\gamma$.
The $s$ parameter may be written in the form:
$s={\it l}_c^2/D^2$, where the length
${\it l}_c=\xi\sqrt{\sigma_{c}/u\sigma_{ab}}$
is the electric field penetration
depth along $z$. For S/I multilayers one has
${\it l}_c\sim \xi_c$ and, as a consequence,
$s\ll 1$ in the temperature range considered in this
paper. For this limit one obtains:
\begin{eqnarray}
\label{small1}
\mu_x^{-1}=[i\omega(M_c+M_{2D})+\eta_c+\eta_{2D}] D^{-1}
\\
\label{small2}
\mu_y^{-1}=\mu_x^{-1}+(i\omega M_j +\eta_j)tan\gamma
\\
\label{small3}
M_{2D}=\frac{\hbar^2 \varepsilon}{8e^2 D}
ln(1+a_d/\xi)\,;\,
\eta_{2D}=\frac{4\pi\sigma_c M_{2D}}{\varepsilon}
\\
\label{small4}
\eta_j\simeq \frac{\sigma_c \phi_0^2}{\pi c^2 DL_j} \,;\,
M_j = \frac{\varepsilon \eta_j}{4\pi\sigma_c}
\end{eqnarray}
where $\eta_c=\eta_0\alpha_1 D$, $M_c=\eta_0\tau\alpha_2 D$ and
$\alpha_1$, $\alpha_2$ are the constants of the order unity.
The terms $\eta_{c}$ and $i\omega M_c$ are connected with the
dissipation and inertial effects in normal core domains.
The logarithmic terms $\eta_{2D}$, $M_{2D}$
in Eqs.~(\ref{small3}) may be considered as the
contributions to the viscosity and inertial mass
of a single 2D pancake
vortex. Let us compare the $M_{2D}$ value to the inertial mass
$M_c$ of the normal core for high-$T_c$ materials
(for the angular domain $tan\gamma\gg \xi/D$).
Considering ${Bi-2:2:1:2}$
as an example we take $D\simeq 15 \AA$,
${\varepsilon\simeq 10}$,
${\sigma_{ab}^{-1}(T\sim T_c)\sim 10^{-4}\Omega cm}$,
${\xi(T=0)=\xi_0\simeq 20-40\AA}$, ${T_c\simeq 80-90K}$,
${L_j/D\simeq 300-1000}$ \cite{muller}
and obtain
$$
\frac{M_{2D}}{M_c}\simeq
\frac{4\varepsilon \xi_0^2 T_c}
{\pi^2 \alpha_2\hbar uD^2\sigma_{ab}}
ln\,\frac{a_d}{\xi} \sim
0.1\, ln\,\frac{a_d}{\xi}
$$
Thus, for $tan\gamma > L_j/D$ one has ${M_{2D}\sim M_c}$.
This rough estimate suggests that the contribution $M_{2D}$
to the vortex
mass of electromagnetic origin may be significant
in layered high-$T_c$ superconductors.
In the angular domain $Dtan\gamma\gg L_j$ the vortex
dynamic mobility depends essentially
on the orientation of the current
with respect to the in-plane magnetic field component.
Such a dependence results
from the contribution to $\mu_y^{-1}$,
proportional to the length $Dtan\gamma$ of the Josephson string
connecting the neighbouring 2D pancake vortices.
The terms $\eta_j$ and $M_j$ correspond to
the viscosity and inertial mass per unit length of
the Josephson vortex.
Note that the expressions (\ref{small4}) are
in good agreement with the ones obtained previously in
Refs.~\cite{clem2,clem1} for vortices parallel to the layers.

The normal currents in nonsuperconducting
layers lead to the penetration
of the field ${\bf E}$ generated by a 2D vortex (moving in the
plane $z=nD$) at a finite length ${\it l}_c$ along $z$.
For S/N multilayers the ${\it l}_c$ value may be larger
than the structure period $D$. This fact results in
an essential decrease of the potential
$\varphi_n$ in the 2D core regions
($\xi<{|{\bf r}-{\bf R}-{\bf r}_{n}| < a_d}$).
As a consequence,
the order parameter phase in these regions
satisfies the diffusion type
equation (see Eq.~(\ref{phase})), where
$D_{\theta}=\xi^2/\tau$ is the diffusion constant.
The characteristic time scale of the
phase distortion propagation through the 2D core region is
$t_0\sim a_d^2/D_{\theta}$.
If the frequency $\omega$ of the applied
ac field is higher than the value $\omega_0\sim t_0^{-1}$, then
the essential time dispersion comes into play and the logarithmic
divergence of the mobility coefficient is cut off at a length
scale $L_{\omega}$ determined by the $\omega$ value.
The upper cutoff $L_{\omega}$ is of the order of a diffusion length
${L_{\omega}= \sqrt{D_{\theta}/\omega}}$.
The validity of these qualitative arguments is proved by direct
calculations.
As an example we consider here only the limit ${{\it l}_c\gg D}$
(${s\gg 1}$) which is relevant to
S/N multilayers, at least, for temperatures close to $T_c$.
We also restrict ourselves to the frequency range
${\omega\tau\ll 1}$ (${L_{\omega}\gg\xi}$).
The complex-valued dynamic mobility has the
form $\mu_{x,y}^{-1}=(1+i\omega \tau/u)\eta_{x,y}$, where
the quantities $M_{x,y}=\tau\eta_{x,y}/u$ and $\eta_{x,y}$
may be considered as the components of the effective
mass and viscosity tensors, respectively.
Evaluating the integral in Eq.~(\ref{equ1}) to the logarithmic
accuracy one obtains:
\begin{eqnarray}
\label{sl1}
\eta_{x}\simeq \eta_0 (F+Q/q);\,
\eta_{y}\simeq \eta_0 (F+qQ)
\\
\label{sl2}
q=\sqrt{1+(Dtan\gamma/L_j)^{2}};\,\,
F=\beta_1 +ln(1+L_m/\xi)\\
\label{sl3}
Q=ln\left(1+\frac{min[L_{\omega},a_d\sqrt{s}]}
{\xi+L_m}\right)
+\beta_2,
\end{eqnarray}
where ${|\beta_{1,2}|\stackrel{_<}{_\sim}1}$,
${L_m=min[a_d,L_{\omega}]}$.
The logarithmic in $\omega$ terms
contribute to the dynamic mobility
as well as to the effective mass and
viscosity in the frequency range
${s^{-1}\xi^2 a_d^{-2}< \omega\tau <1}$.
For ${\sqrt{s}Dtan\gamma<\xi}$
the logarithmic terms in Eqs.~(\ref{sl2}),(\ref{sl3})
are small and the effective mass and viscosity are
determined by the dynamic processes in the normal core regions.
In the opposite limit (${\sqrt{s}Dtan\gamma>\xi}$)
these terms are large and the values $\beta_{1}$, $\beta_{2}$
may be neglected in a wide frequency range.
The latter conclusion is not valid only for
${Dtan\gamma\gg L_{j}}$ and ${L_{\omega}\ll L_j}$,
when the logarithmic term in Eq.~(\ref{sl3}) vanishes
while the value ${\beta_2\sim -iL_{\omega}^2/L_j^2}$
provides an essential imaginary contribution
to the coefficient $\eta_y$.
As a consequence, the $M_y$ and $\eta_y$ values introduced above
become complex-valued and, hence, in this case
it is not adequate to define these
quantities as the effective mass and viscosity
(otherwise these definitions may be used).
For higher frequences ${\omega\tau>1}$
the vortex mobility is completely determined
by the order parameter dynamics in the normal core region, which
was analysed, e.g., in Ref.~\cite{kupr}.

In summary, we have obtained the
equation of tilted vortex motion,
which may be applicable to various layered systems, including
high-$T_c$ copper oxides.
The angular dependences of the viscosity coefficient
and the inertial mass have been investigated.
For S/N mutilayers the complicated effective core
structure is shown to result
in the specific frequency dependence  of
the complex-valued dynamic mobility even in the frequency range
${\omega\tau<1}$.
The approach developed above provides a starting point
for the study of high-frequency linear response
in the mixed state of layered superconductors.
For this purpose the vortex dynamic mobility should be certainly
extended to include both pinning and flux-creep effects
\cite{mobil,clem2}.

This work was supported, in part, by a fellowship of the
International Center for Fundumental Physics in Moscow,
Russian Foundation for Fundumental Research
(Grant No. 96-02-16993a)
and Russian State Program on Condensed Matter Physics
(Grant No.95042).
%
%

\end{document}